\documentclass[prd,10pt,showpacs,preprintnumbers,amsmath,amssymb,twocolumn]{revtex4-2}

\usepackage{graphicx}
\usepackage{dcolumn}
\usepackage{bm}
\usepackage{amssymb}
\usepackage{amsmath}
\usepackage{amsthm}
\usepackage{yhmath}
\usepackage{slashed}
\usepackage{xcolor}

\def\bmA{{\bm A}}
\def\bmB{{\bm B}}
\def\bmC{{\bm C}}

\newcounter{mnotecount}

\newcommand{\mnotex}[1]
{\protect{\stepcounter{mnotecount}}$^{\mbox{\footnotesize $\bullet$\themnotecount}}$ 
\marginpar{
\raggedright\tiny\em
$\!\!\!\!\!\!\,\bullet$\themnotecount: #1} }

\begin{document}

\preprint{}

\title{New spinorial approach to mass inequalities for black holes in general relativity}

\author{Jaros\l aw Kopi\'nski}
 \email{jkopinski@cft.edu.pl}
\affiliation{Faculty of Physics, University of Warsaw, Pasteura 5, 02-093 Warsaw, Poland}
\author{Juan A. Valiente Kroon}%
 \email{j.a.valiente-kroon@qmul.ac.uk}
\affiliation{%
School of Mathematical Sciences, Queen Mary University of London
Mile End Road, London E1 4NS, UK
}%

\date{\today}

\begin{abstract}
\noindent
A new spinorial strategy for the construction of geometric inequalities involving the Arnowitt-Deser-Misner (ADM) mass of black hole systems in general relativity is presented. This approach is based on a second order elliptic equation (the approximate twistor equation) for a valence 1 Weyl spinor. This has the advantage over other spinorial approaches to the construction of geometric inequalities based on the Sen-Witten-Dirac equation that it allows to specify boundary conditions for the two components of the spinor. This greater control on the boundary data has the potential of giving rise to new geometric inequalities involving the mass. In particular, it is shown that the mass is bounded from below by an integral functional over a marginally outer trapped surface (MOTS) which depends on a freely specifiable valence 1 spinor. From this main inequality, by choosing the free data in an appropriate way, one obtains a new nontrivial bounds of the mass in terms of the inner expansion of the MOTS. The analysis makes use of a new formalism for the $1+1+2$ decomposition of spinorial equations.
\end{abstract}

\maketitle

\noindent
\section{Introduction} 
Geometric inequalities are a prime example of the rich interplay between general relativity and geometric analysis. They relate quantities with a clear physical content with geometric structures of the spacetime. In particular, they provide important qualitative insight into fundamental aspects of black holes. 

The most fundamental geometric inequality in general relativity is, without doubt, the so-called \emph{positivity of the ADM mass}. Although a proof of this result (for axially symmetric spacetimes) can be found in the work of Brill \cite{Bri59}, a first \emph{general} proof was obtained by Schoen and Yau \cite{SchYau79,SchYau81a} using methods of geometric analysis.  An alternative proof, using spinorial methods, was later
given by Witten \cite{Wit81}. An extension of this last result, showing the positivity of the mass for black hole spacetimes was given in \cite{GibHawHorPer83}. Technical aspects of the spinorial proof, including the existence of solutions of the boundary value problem for the Sen-Witten-Dirac equation have been addressed in \cite{ReuTod84,Her98a}. A further refinement of the positivity of the mass is given the so-called \emph{Penrose inequality}. It provides a lower bound of the mass of a black hole in terms of (the square root of)  its area---see e.g. \cite{Pen73}---and is closely related to the \emph{Cosmic Censorship conjecture}. The Penrose inequality has only been rigorously proved in the so-called \emph{Riemannian case} (i.e. when the initial hypersurface is time symmetric)---see \cite{HuiIlm01}, also \cite{Mar09} for a survey on the subject. This proof makes use of powerful methods of geometric analysis to study the properties of a geometric flow. In the case of axisymmetric black holes, alternative bounds for the mass in terms of the angular momentum (mass-angular momentum inequalities) have been analysed and rigorously proven \cite{Dai06b,Dai08}---see also \cite{DaiGab18} for a review on the subject.

The proof of the positivity of the mass for black holes in \cite{GibHawHorPer83} suggests that it may be possible to make use of (an extension of) Witten's strategy to obtain non-trivial bounds on the mass and, in particular, obtain a proof the general Penrose inequality. Indeed, a Penrose-like inequality has been obtained in \cite{Her97b} by this approach---however, the \emph{classical} Penrose inequality remains, so far, unproven. One of the main advantages of the spinorial approach to the construction of geometric inequalities is that it leads to conceptually clearer arguments. For a 4-dimensional spacetime, the existence of a spin structure does not introduce any additional restrictions, so working in the setting of  asymptotically flat (or Schwarzschildean) hypersurfaces, one can obtain bounds on the mass directly from the existence of solution of a certain spinorial equation. However, the resulting inequality will depend heavily on the boundary conditions. 

Witten's argument for the positivity of the mass makes use of an integral identity for a spinor field $\kappa_A$ over a 3-dimensional hypersurface $\mathcal{S}$. This identity contains both bulk and boundary integrals. One part of the bulk integrals is manifestly non-negative while the rest can be eliminated if $\kappa_A$ satisfies the Sen-Witten-Dirac equation $\mathcal{D}_A{}^B\kappa_B=0$ (see below for an explanation of the notation). If suitable asymptotic conditions for $\kappa_A$ are prescribed then the boundary integral at infinity can be shown to be related to the mass. Thus, in order to obtain a non-trivial bound on the latter one is left with the task of identifying conditions on the inner (black hole) boundary which ensure the solvability of the Sen-Witten-Dirac equation and such that the inner boundary integral is non-negative---e.g. it involves the area. As the analyses in \cite{GibHawHorPer83,ReuTod84,Her98} show, a limitation of Witten's strategy is that the  Sen-Witten-Dirac equation is first order elliptic and thus, roughly speaking one can only prescribe one of the components of $\kappa_A$.

\smallskip
\noindent
\textbf{Main results.} In this paper we develop a different spinorial framework for the study
of geometric inequalities involving the ADM mass which addresses the
difficulties in Witten's approach of prescribing boundary data. This
strategy builds on the analysis of the so-called \emph{approximate
  twistor equation} introduced in \cite{BaeVal11a}. The approximate
twistor equation is a second order elliptic equation for a Weyl spinor
$\kappa_A$ on a 3-dimensional asymptotically Euclidean manifold, which
is assumed to be a hypersurface of a vacuum spacetime
$(\mathcal{M}, g_{ab})$. Using suitably constructed solutions to the
boundary value problem for the approximate twistor equation we find
that for a MOTS one has the inequality
\[
4\pi m \geq \frac{\kappa}{\sqrt{2}} \mathtt{H}[\phi_A,\bar\phi_{A'}],
\qquad \kappa \equiv 8\pi G/c^4,
\]
where $m$ denotes the ADM mass of the spacetime $(\mathcal{M},g_{ab})$
and
$\mathtt{H}[\phi_A,\bar\phi_{A'}]$ is the Nester-Witten functional
over the MOTS evaluated on a freely specifiable spinor $\phi_A$ over
the 2-surface. This \emph{master inequality} can be used as the
starting point for the systematic construction of geometric
inequalities involving the mass. In particular, a new proof of the
positivity of the mass for black holes follows directly from the above
inequality. A couple of further examples of inequalities which follow
directly from the master inequality are provided in the main text. 

\medskip
A substantial part of the calculations in this article have been
carried out in the suite of packages {\tt xAct} for tensor and spinor
manipulations in Mathematica \cite{xAct}. In particular, we have profited from the
package {\tt SpinFrames} allowing computations in the NP and GHP formalisms.

\medskip
\noindent
\textbf{\em Organisation of the article.} This paper is organized as
follows. In Sec. II we establish the framework of $1+1+2$
space--spinor formalism in which we are working. Next section is
dedicated to the approximate twistor equation, which together with the
appropriate boundary condition will be used in Sec. IV to establish a
new bound on the ADM mass of the initial data. The role of appendices
is to clarify the arguments used in the main body of the paper.

\medskip
\noindent
\textbf{\em Notation and conventions.} In the following, 4-dimensional
metrics are taken to have signature $(+---)$. Consequently, Riemannian
3- and 2-dimensional metrics are taken to be negative definite. When
convenient, we expand spinorial expressions using the
Geroch-Held-Penrose (GHP) formalism. In using spinors and the GHP
formalism, we follow the conventions of \cite{PenRin84}. The Einstein
field equations are given by $G_{ab} = \kappa T_{ab}$ where as usual
$\kappa \equiv 8\pi G/c^4$.

\section{The $1+1+2$ space--spinor formalism}
\label{Appendix:OnePlusOnePlus2}
Consider initial data sets
 $(\mathcal{S},h_{ij},K_{ij})$ for the vacuum Einstein field equations satisfying in the asymptotic region the conditions
\begin{subequations}
\begin{eqnarray}
&& h_{ij} = - \left(1+\frac{2m}{r}\right)\delta_{ij} + o_\infty(r^{-3/2}),\label{DecayAssumption1} \\
&& K_{ij} = o_\infty(r^{-5/2}), \label{DecayAssumption2}
\end{eqnarray}
\end{subequations}
 with $r^2\equiv (x^1)^2 + (x^2)^2 + (x^3)^2$, $(x^1,x^2,x^3)$
 asymptotically Cartesian coordinates and $m$ the ADM mass. Initial data sets of this type are called
 \emph{asymptotically Schwarzschildean}. In addition, it is assumed that $\mathcal{S}$ has one inner boundary $\partial \mathcal{S}\approx \mathbb{S}^2$. 
 
The $1+1+2$ spinor formalism is inspired by the 2-dimensional Sen connection introduced in \cite{Sza94a} which uses $SL(2,\mathbb{C})$ spinors. Here we adapt these ideas to $SU(2,\mathbb{C})$ spinors (the so-called space spinors first introduced in \cite{Som80}) which allows to work only with spinors with unprimmed indices. A discussion of the space spinor formalism can be found in \cite{CFEBook}---see also \cite{Ash91}. 

\subsection{Basic setting}
Let $\tau^{AA'}$ and $\rho^{AA'}$ denote, respectively the
spinorial counterpart of the 
(timelike) normal to the hypersurface $\mathcal{S}$ and the
(spacelike) normal to $\partial \mathcal{S}$ on $\mathcal{S}$. We
consider spinor dyads $\{ o^A,\, \iota^A \}$ such that
\begin{eqnarray*}
&& \tau_{AA'} \tau^{BA'} = \delta_A{}^B \Longrightarrow
   \tau_{AA'}\tau^{AA'} =2, \\
&& \rho_{AA'}\rho^{BA'} = -\delta_A{}^B \Longrightarrow
   \rho_{AA'}\rho^{AA'} =-2.
\end{eqnarray*}
The spinors $\tau^{AA'}$ and $\rho^{AA'}$ are Hermitian. 
We require $\tau^{AA'}$ and $\rho^{AA'}$ to be orthogonal to each
other---that is, $\tau_{AA'}\rho^{AA'} =0$. The
\emph{complex metric} can now be defined as $\gamma_{AB}\equiv \tau_B{}^{A'}
\rho_{AA'}$. It follows from the definition that
\[
\gamma_A{}^B\gamma_B{}^C =\delta_A{}^C. 
\] 
Because of the orthogonality of $\tau^{AA'}$ and $\rho^{AA'}$ the complex metric is a symmetric spinor, $\gamma_{AB}=\gamma_{(AB)}$. 

\smallskip
The projector to the
2-dimensional surface $\partial\mathcal{S}$ admits the alternative
expressions
\begin{eqnarray*}
&& \Pi_{AA'}{}^{BB'} = P_{AA'}{}^{QQ'} T_{QQ'}{}^{BB'} \\
&&\phantom{\Pi_{AA'}{}^{BB'}} = \delta_A{}^B \delta_{A'}{}^{B'}
   -\tfrac{1}{2}\tau_{AA'}\tau^{BB'} +
   \tfrac{1}{2}\rho_{AA'}\rho^{BB'}\\
  &&\phantom{\Pi_{AA'}{}^{BB'}} = \tfrac{1}{2}( \delta_A{}^B
       \delta_{A'}{}^{B'} -\gamma_A{}^B\bar{\gamma}_{A'}{}^{B'}),
\end{eqnarray*}
where
\begin{eqnarray*}
&& P_{AA'}{}^{BB'} \equiv \delta_A{}^B \delta_{A'}{}^{B'} +
   \frac{1}{2}\rho_{AA'}\rho^{BB'}, \\
&& T_{AA'}{}^{BB'}\equiv \delta_A{}^B \delta_{A'}{}^{B'}
   -\frac{1}{2}\tau_{AA'}\tau^{BB'} 
\end{eqnarray*}
denote, respectively, the projectors to the distributions generated by
$\rho^{AA'}$ and $\tau^{AA'}$.

\smallskip
Several of the calculations 
simplify if one makes use of an adapted spin dyad $\{ o_A,\, \iota_A
\}$ with $o_A\iota^A=1$ such that $\hat{o}_A=\iota_A$ and $\hat{\iota}_A =-o_A$, where $\string^$ denotes the Hermitian conjugation. We have
\[
\tau^{AA'} = o^A \bar{o}^{A'} +\iota^A \bar{\iota}^{A'}.
\]
It then follows that
\begin{eqnarray*}
&& \rho^{AA'} = o^A \bar{o}^{A'} -\iota^A \bar{\iota}^{A'}, \\
&& \gamma_{AB} =o_A \iota_B + o_B\iota_A.
\end{eqnarray*}
The above construction, restricted to the 2-dimensional surface
$\partial \mathcal{S}$ still allows the freedom of a rotation
\[
o^A \mapsto e^{i \vartheta} o^A, \qquad \iota^A \mapsto e^{-i
  \vartheta} \iota^A.
\]
If one defines, following standard conventions, components of a spinor $\kappa_A$ with respect to $\{ o_A,\, \iota_A
\}$ by
\[
\kappa_0 \equiv o^A\kappa_A, \qquad \kappa_1\equiv \iota^A\kappa_A,
\]
then
\[
\hat{\kappa}_A = \overline{\kappa}_0 o_A + \overline{\kappa}_1 \iota_A. 
\]

\subsection{The 3-dimensional and 2-dimensional Sen connections}  The \emph{3-dimensional} and
\emph{2-dimensional} Sen connections are defined, respectively, by
\begin{subequations}
\begin{eqnarray}
&& \mathcal{D}_{AA'} \kappa_C \equiv T_{AA'}{}^{BB'} \nabla_{BB'}
   \kappa_C, \label{SLSen3D}\\
&& \slashed{\mathcal{D}}_{AA'} \kappa_C \equiv \Pi_{AA'}{}^{BB'} \nabla_{BB'}
   \kappa_C. \label{SLSen2D}
\end{eqnarray}
\end{subequations}
One can use the spinor $\tau^{AA'}$ to obtain $SU(2,\mathbb{C})$
(i.e. \emph{space spinor}) versions of the the  above derivatives. More precisely, one has 
\[
\mathcal{D}_{AB} \equiv \tau_{(B}{}^{A'} \mathcal{D}_{A)A'}, \qquad
\slashed{\mathcal{D}}_{AB} \equiv \tau_{(B}{}^{A'}
\slashed{\mathcal{D}}_{A)A'}. 
\]
From the above expressions one can derive the following alternative
expressions:
\[
\mathcal{D}_{AB}\kappa_C \equiv \tau_{(B}{}^{A'} \nabla_{A)A'}
\kappa_C, \qquad \slashed{\mathcal{D}}_{AB} \equiv \gamma_B{}^Q\gamma_{(A}{}^P
\mathcal{D}_{Q)P}. 
\] 
Moreover, one has the decompositions
\begin{eqnarray*}
&& \nabla_{AA'} = \tfrac{1}{2}\tau_{AA'} \mathcal{P} -\tau^Q{}_{A'}
   \mathcal{D}_{AQ},  \label{DecompositionDerivatives1}\\
&& \mathcal{D}_{AB} = \slashed{\mathcal{D}}_{AB} -
   \tfrac{1}{2}\gamma_{AB} \slashed{\mathcal{D}}, \label{DecompositionDerivatives2}
\end{eqnarray*}
where
\[
\mathcal{P}\equiv \tau^{AA'} \nabla_{AA'}, \qquad
\slashed{\mathcal{D}}\equiv \gamma^{AB} \mathcal{D}_{AB}
\]
are directional derivatives in the direction of $\tau_{AA'}$ and
$\gamma_{AB}$, respectively. 

\subsection{The extrinsic curvature}
Following the standard definition adapted to the present setting, the Weingarten spinor associated with generator $\tau^{AA'}$ is given by
\[
K_{ABCD} =  \tau_D{}^{C'} \mathcal{D}_{AB} \tau_{CC'}
\]
We will assume that distribution is integrable, i.e. $K_{ABCD}$ corresponds to the extrinsic curvature of a hypersurface orthogonal to $\tau^{AA'}$. This is equivalent to the condition
\[
K_{AC}{}^C{}_B= \tfrac{1}{2}K\epsilon_{AB},
\]
where $K=K_{AB}{}^{AB}$ is the mean curvature of $\mathcal{S}$ and $\epsilon_{AB}$ is the antisymmetric spinor generating symplectic bilinear form. It will also be convenient to introduce a complete symmetrisation of extrinsic curvature, $\Omega_{ABCD}$. It can be defined by the following relation
\[
K_{ABCD} =\Omega_{ABCD} - \tfrac{1}{3}K \epsilon_{A(C}\epsilon_{D)B}.
\]
\subsection{Levi-Civita connections} 
The spinor form of
the induced metric $h_{ij}$ on $\mathcal{S}$ can be obtained from the projector
$T_{AA'}{}^{BB'}$ by removing primed indices using the spinor
$\tau_{AA'}$. After using the Jacobi identity for $\epsilon_{AB}$ one
finds that
\[
h_{ABCD} \equiv -\epsilon_{A(C}\epsilon_{D)B}.
\]
One can verify that
\begin{eqnarray*}
&& h_{ABCD}= h_{CDAB}, \\
&&  h_{ABCD}=h_{(AB)CD} =h_{AB(CD)}=h_{(AB)(CD)}.
\end{eqnarray*}
Similarly, from $\Pi_{AA'}{}^{BB'}$ a calculation readily gives the
expression
\[
\sigma_{ABCD} = \tfrac{1}{2}(\epsilon_{AC}\epsilon_{BD} + \gamma_{AC}\gamma_{BD})
\]
for the induced metric $\sigma_{ab}$ on $\partial \mathcal{S}$. To obtain
this last expression it has been used that
$\hat{\gamma}_{AB}=-\gamma_{AB}$. 

\smallskip
Let $D_{AB}$ and $\slashed{D}_{AB}$ denote, respectively, the
$SU(2,\mathbb{C})$ form of the Levi-Civita connection of the metrics
$h_{ab}$ and $\sigma_{ab}$. One has that 
\[
D_{AB} \epsilon_{CD}=0, \qquad \slashed{D}_{AB} \epsilon_{CD}=0.
\]
In addition,
\[
\slashed{D}_{AB} \gamma_{CD}=0.
\]
The relation between the Sen and Levi-Civita connections can be worked
out using the \emph{standard} tricks---see e.g. \cite{PenRin84}. One finds that 
\begin{eqnarray*}
&& \mathcal{D}_{AB}\pi_C = D_{AB}\pi_C + \tfrac{1}{2}K_{ABC}{}^Q\pi_Q,
  \\
&& \slashed{\mathcal{D}}_{AB} \pi_C = \slashed{D}_{AB}\pi_C + Q_{AB}{}^Q{}_C\pi_Q,
\end{eqnarray*}
where, for convenience, we have defined the \emph{transition spinor}
\[
Q_{AB}{}^C{}_D \equiv -\tfrac{1}{2}\gamma_D{}^Q \slashed{\mathcal{D}}_{AB}\gamma_Q{}^C.
\]
Using the GHP formalism \cite{GerHelPen73,PenRin84} one can arrive at
\begin{eqnarray*}
&& Q_{ABCD}= \sigma' o_A o_B o_C o_D + \sigma \iota_A\iota_B\iota_C\iota_D \\
&& \hspace{2cm} - \rho o_A o_B \iota_C \iota_D - \rho' \iota_A \iota_B o_C o_D.
\end{eqnarray*}

The Levi-Civita covariant derivatives are real in the sense that
\[
\widehat{D_{AB} \pi_C} = - D_{AB}\hat{\pi}_C, \qquad \widehat{\slashed{D}_{AB} \pi_C} = - \slashed{D}_{AB}\hat{\pi}_C.
\]
This implies the following formulas for Hermitian conjugation of Sen derivatives:
\begin{eqnarray*}
&& \widehat{\mathcal{D}_{AB} \pi_C} = -\mathcal{D}_{AB} \hat{\pi}_C +
   K_{ABC}{}^D \hat{\pi}_D, \\
&& \widehat{\slashed{\mathcal{D}}_{AB} \pi_C}= -\slashed{\mathcal{D}}_{AB}\hat{\pi}_C
   + \left(Q_{ABC}{}^D + \hat{Q}_{ABC}{}^D \right) \hat{\pi}_D.
\end{eqnarray*}
Finally observe that a direct computation gives
\begin{eqnarray*}
&& \slashed{D}_{AB}o_C = \alpha o_A o_B o_C -\beta \iota_A\iota_B o_C,\\
&& \slashed{D}_{AB}\iota_C = \beta \iota_A\iota_B\iota_C -\alpha o_A o_B \iota_C.
\end{eqnarray*}
However, computing the Hermitian conjugate of the first expression one readily has that
\[
\slashed{D}_{AB}\iota_C =-\bar{\alpha} \iota_A\iota_B\iota_C +\bar\beta o_A o_B \iota_C.
\]
Hence, one concludes that
\[
\alpha +\bar\beta =0,
\]
This relation leads to the  formula
\[
 \left( \slashed{D}_{AC} \slashed{D}_{B}{}^C - \slashed{D}_{B}{}^C \slashed{D}_{AC} \right) \kappa^B = \left( \rho \rho' - \sigma \sigma' + \Psi_2 \right) \kappa_A,
\]
satisfied in the vacuum spacetime.

It is worth noticing that the condition $\alpha + \bar \beta = 0$, which needs to be imposed to ensure the reality of $\slashed{D}_{AB}$, implies that this connection is generally not Levi-Civita. However, since it is introduced here in a similar natural way to the 3-dimensional Levi-Civita connection (compare the formulas for the transition spinors $Q_{ABCD}$ and $K_{ABCD}$), we will assume that $\alpha + \bar \beta = 0$ holds on the boundary $\partial S$ and treat it as a gauge condition. By doing so, we can consider $\slashed{D}_{AB}$ as a Levi-Civita connection. Further discussion on this matter will be moved elsewhere.

\subsection{MOTS}
Let $l^a$ and $k^a$ denote
future-oriented null vectors spanning the normal bundle to $\partial
\mathcal{S}$ and such that $l^a k_a=1$. The expansions associated to
$l^a$ and $k^a$ are defined, respectively, by
\[
\theta^+ \equiv \sigma^{ab}\nabla_a l_b, \qquad \theta^-\equiv \sigma^{ab}
\nabla_a k_b.
\]
Our conventions are that $l^a$ denotes an \emph{outgoing} null vector
whereas $k^a$ is an \emph{ingoing} one. The 2-surface $\partial
\mathcal{S}$ is said to be a MOTS if
$\theta^+=0$ and $\theta^-\leq 0$. Let $l^{AA'}$ and $k^{AA'}$ denote the spinorial
counterparts of $l^a$ and $k^a$. A natural choice for $l^a$ and $k^a$
is given by
\[
l^a=\tfrac{1}{2}( \tau^a + \rho^a), \qquad k^a =\tfrac{1}{2}(\tau^a -\rho^a),
\]
so that 
\[
l^{AA'} = o^A\bar{o}^{A'}, \qquad k^{AA'}=\iota^A \bar{\iota}^{A'}. 
\]
A computation then shows that in terms of the GHP formalism one has
that
\[
\theta^+= -\rho-\bar{\rho}, \qquad \theta^-=-\rho'-\bar{\rho}'.
\]
In the present setting one has, moreover, that both $\rho$ and
$\rho'$ are real (see \cite{PenRin84}, Proposition 4.14.2) so that, in fact, one has that 
\begin{equation} \label{thrho}
\theta^+ =-2\rho, \qquad \theta^-=-2\rho'.
\end{equation}
The contraction $Q_A{}^P{}_{CP}$ will play an important role in the sequel. An expansion in terms of the dyad readily shows that
\[
Q_A{}^P{}_{BP} = \rho\, o_A\iota_B -\rho'\, \iota_A o_B, 
\] 
If $\rho$ and $\rho'$ are real, then it
readily follows that 
\[
\hat{Q}_A{}^P{}_{BP} = -\big( \rho\, o_B\iota_A -\rho'\, \iota_B o_A
\big) =-Q_B{}^P{}_{AP}. 
\]
Observing that 
\[
o_A\iota_B = \frac{1}{2}\gamma_{AB} + \frac{1}{2}\epsilon_{AB}
\]
one obtains the more convenient expression
\[
Q_{A}{}^C{}_{BC} =\frac{1}{2}(\rho-\rho') \gamma_{AB} + \frac{1}{2}(\rho+\rho') \epsilon_{AB}.
\]
In particular, for a MOTS one has
\[
Q_{A}{}^C{}_{BC} =\frac{1}{2}\rho' \left(   \epsilon_{AB} -\gamma_{AB} \right). 
\]
\section{The approximate twistor equation}
Let $\mathfrak{S}_1$, $\mathfrak{S}_3$ denote, respectively, the spaces of valence 1 and 3 symmetric spinors over the hypersurface $\mathcal{S}$. One defines the \emph{spatial twistor operator} 
\[
\mathbf{T}: \mathfrak{S}_1\rightarrow \mathfrak{S}_3, \qquad
\mathbf{T}(\kappa)_{ABC} = \mathcal{D}_{(AB}\kappa_{C)}. 
\]
The operator $\mathbf{T}$ can be easily shown to be overdetermined elliptic. The equation $\mathcal{D}_{(AB}\kappa_{C)}=0$ arises from the space-spinor decomposition of the twistor equation $\nabla_{A'(A}\kappa_{B)}=0$ \cite{BaeVal11a}. 
The formal adjoint of $\mathbf{T}$, to be denoted by $\mathbf{T}^*$, is given by 
\[
 \mathbf{T}^*: \mathfrak{S}_3\rightarrow \mathfrak{S}_1, 
\qquad \mathbf{T}^*(\zeta)_A
 \equiv \mathcal{D}^{BC}\zeta_{ABC} - \Omega_A{}^{BCD}\zeta_{BCD}.
\]
The operator $\mathbf{T}^*$ can be shown to be underdetermined elliptic. The \emph{approximate twistor equation} follows from considering the composition operator $\mathbf{L}\equiv
\mathbf{T}^*\circ\mathbf{T}:\mathfrak{S}_1\rightarrow\mathfrak{S}_1$ and is given by 
\begin{equation}
\mathbf{L}(\kappa_A) \equiv
\mathcal{D}^{BC}\mathcal{D}_{(AB}\kappa_{C)}
-\Omega_A{}^{BCD}\mathcal{D}_{BC}\kappa_D=0.
\label{ApproximateTwistorEquation}
\end{equation}
By construction the operator given by equation
\eqref{ApproximateTwistorEquation} is formally self-adjoint elliptic---i.e. $\mathbf{L}^*=\mathbf{L
}$. Given a solution $\kappa_A$ to equation \eqref{ApproximateTwistorEquation}, it is convenient to define the spinors $\xi_A \equiv \tfrac{2}{3}\mathcal{D}_A{}^Q\kappa_Q$ and $\xi_{ABC}\equiv \mathcal{D}_{(AB}\kappa_{C)}$ encoding the independent components of the derivative $\mathcal{D}_{AB}\kappa_C$. Moreover, set $\zeta_A \equiv \hat{\xi}_A$. A key observation is the following: if $\kappa_A$ satisfies
$\mathbf{L}(\kappa_A)=0$, then using the properties of the Hermitian
conjugation one has that $\mathbf{L}(\zeta_A)=0$.

\medskip
In the following we consider 
solutions to equation \eqref{ApproximateTwistorEquation} with an
asymptotic behaviour of the form
\begin{equation}
\kappa_\mathbf{A} =\bigg( 1+\frac{m}{r} \bigg) x_{\mathbf{A}\mathbf{B}} o^{\mathbf{B}} +
o_\infty(r^{-1/2}) 
\label{AsymptoticBehaviourSpinor}
\end{equation}
where given some asymptotically Cartesian coordinates
$\underline{x}=(x^\alpha) $ we set
\[
x_{\mathbf{A}\mathbf{B}} \equiv \frac{1}{\sqrt{2}}
\left( 
\begin{array}{cc}
x^1+\mbox{i}x^2 & -x^3 \\
-x^3 & -x^1+\mbox{i} x^2
\end{array}
\right)
\]
and the spinor $o^A$ is part of a normalised spin dyad $\{o^A,\, \iota^A\}$
adapted to $\mathcal{S}$---that is, $\iota^A=\hat{o}^A$.  A computation reveals that
\begin{subequations}
\begin{eqnarray}
&& \xi_{\mathbf{A}} =\bigg (1 -\frac{m}{r}\bigg) o_{\mathbf{A}} + o_{\infty}(r^{-3/2}), \label{DecayXi1}\\
&& \mathcal{D}_{(\mathbf{AB}}\xi_{\mathbf{C})} = -\frac{3 m}{2 r^3} x_{(\mathbf{AB}}
   o_{\mathbf{C})} + o_\infty(r^{-5/2}). \label{DecayXi2}
\end{eqnarray}
\end{subequations}

\smallskip
\noindent
\subsection{Relation to the ADM mass} Central to our analysis is the functional
\[
\mathtt{I}[\kappa_A] \equiv \int_{\mathcal{S}} \mathcal{D}_{(AB} \zeta_{C)}
\widehat{\mathcal{D}^{AB}\zeta{}^{C}} \mbox{d} \mu\geq 0, 
\]
first considered in \cite{BaeVal11a}. If $\mathbf{L}(\kappa_A)=0$ then integrating by parts it is possible to rewrite $\mathtt{I}[\kappa_A]$ in terms of boundary integrals at the sphere at infinity
$(\partial\mathcal{S}_\infty)$ and the inner boundary ($\partial\mathcal{S}$):
\[
\mathtt{I}[\kappa_A]= \oint_{\partial \mathcal{S}_\infty } n_{AB} \zeta_C \widehat{\mathcal{D}^{(AB} \zeta^{C)}} \mbox{d} S - \oint_{\partial \mathcal{S} } n_{AB} \zeta_C \widehat{\mathcal{D}^{(AB} \zeta^{C)}} \mbox{d} S.
\]
As
a consequence of the asymptotic expansions \eqref{DecayXi1}-\eqref{DecayXi2} the integral over $\partial\mathcal{S}_\infty$ can be shown to equal $4\pi m$. Thus, it follows that 
\begin{equation}
4\pi m \geq \oint_{\partial \mathcal{S} } n_{AB} \zeta_C \widehat{\mathcal{D}^{(AB} \zeta^{C)}} \mbox{d} S. 
\label{BasicInequality}
\end{equation}

\smallskip
\noindent
\subsection{A boundary value problem} The
inequality \eqref{BasicInequality} suggests considering boundary
conditions of the form $\zeta_A = \phi_A$ where $\phi_A$ is a smooth, freely specifiable 
spinorial field over $\partial\mathcal{S}$. Written in terms of $\kappa_A$ one obtains the condition 
\begin{equation}
\mathcal{D}_A{}^Q \kappa_Q = -\tfrac{3}{2}\hat{\phi}_A, \qquad \mbox{on}
\quad \partial \mathcal{S}. \label{BoundaryCondition}
\end{equation}
The approximate twistor equation  together with the above \emph{transverse boundary condition} can be shown to satisfy the \emph{Lopatinskij-Shapiro compatibility conditions}---see e.g. \cite{Dai06,WloRowLaw95}.   It follows that
the boundary value problem over $\mathcal{S}$ given by \eqref{ApproximateTwistorEquation}
and \eqref{BoundaryCondition} is elliptic. In the following we consider solutions to the associated boundary
value problem with the asymptotic behaviour
\eqref{AsymptoticBehaviourSpinor} and the Ansatz
\begin{equation}
\kappa_A =\mathring{\kappa}_A + \theta_A, \qquad \theta_A \in
H^2_{-1/2}
\label{Ansatz:SolutionApproximateTwistor}
\end{equation}
with $\mathring{\kappa}_A$ given by the leading term in
\eqref{AsymptoticBehaviourSpinor} and where $H^s_\beta$ with $s\in\mathbb{Z}^+$ and $\beta\in \mathbb{R}$ denotes the weighted $L^2$ Sobolev spaces. We follow the conventions for
  these spaces set in \cite{Bar86}. In view of the decay conditions
\eqref{DecayAssumption1}-\eqref{DecayAssumption2}  the
elliptic operator $\mathbf{L}$ is \emph{asymptotically homogeneous}
---see \cite{Can81,Loc81}. This is the standard assumption on elliptic
operators on asymptotically Euclidean manifolds. 

\subsection{ Solvability of the boundary value problem}
To discuss the solvability of the
approximate twistor equation we need to consider \emph{Green's
  identity} for the approximate twistor operator $\mathbf{L}$. That is,
\begin{eqnarray*}
&&\hspace{-5mm} \int_{\mathcal{S}} \mathbf{L}(\kappa_A) \hat{\pi}^A \mathrm{d}\mu - \int_{\mathcal{S}}
\kappa_A \widehat{\mathbf{L}(\pi^A)} \mathrm{d}\mu \\
&& \hspace{5mm} =
\oint_{\partial\mathcal{S}}\left( \mathcal{D}_{(AB} \kappa_{C)}
   n^{AB}\hat{\pi}^C - n^{AB} \kappa^{C} \widehat{\mathcal{D}_{(AB} \pi_{C)}}\right) \mathrm{d}S,
\end{eqnarray*}
where in the above expression it has explicitly been used that
$\mathbf{L}$ is self-adjoint. The
first task is to rewrite the boundary conditions in terms of the
boundary operator $\mathcal{D}_A{}^Q\kappa_Q$ so that one can identify
the natural adjoint boundary conditions. One aims for an identity of the form
\begin{eqnarray*}
&&\hspace{-5mm} \int_{\mathcal{S}} \mathbf{L}(\kappa_A) \hat{\pi}^A \mathrm{d}\mu - \int_{\mathcal{S}}
\kappa^A \widehat{\mathbf{L}(\pi_A)} \mathrm{d}\mu \\
&& \hspace{5mm} =
\oint_{\partial\mathcal{S}}\left( \mathbf{B}(\kappa_A) \hat{\pi}^A -
   \kappa_A \widehat{\mathbf{B}^*(\pi^A)}\right) \mathrm{d}S,
\end{eqnarray*}
where $\mathbf{B}$ is some \emph{natural boundary operator} yet to be identified and $\mathbf{B}^*$ is its formal adjoint over $\partial\mathcal{S}$. Now, the
decomposition of the 3-dimensional Sen connection yields 
\[
\sqrt{2} \mathcal{D}_{(AB}\kappa_{C)} n^{AB} \hat{\pi}^C =
\slashed{\mathcal{D}} \kappa_C \hat{\pi}^C + \xi_A \gamma^A{}_C \hat{\pi}^C.
\]
A further computation shows that the normal derivative
$\slashed{\mathcal{D}}\kappa_C$ can be expressed in terms of $\xi_A$ and
the intrinsic derivative $\slashed{\mathcal{D}}_A{}^Q\kappa_Q$ as
\[
\slashed{\mathcal{D}} \kappa_C = 2 \gamma_C{}^P
\slashed{\mathcal{D}}^Q{}_P\kappa_Q -3 \gamma_C{}^Q \xi_Q. 
\] 
Combining the above expressions one obtains
\[
\mathcal{D}_{(AB} \kappa_{C)} n^{AB} \hat{\pi}^C = \sqrt{2} \left(
  \gamma_C{}^P \slashed{\mathcal{D}}^Q{}_P\kappa_Q \hat{\pi}^C -\gamma_C{}^P \xi_P
  \hat{\pi}^C \right). 
\]
For convenience, define the boundary operator
\[
\mathbf{B}(\kappa_A) \equiv -\sqrt{2} \gamma_A{}^P \xi_P =
-\tfrac{2\sqrt{2}}{3}\gamma_A{}^P \mathcal{D}^Q{}_P\kappa_Q.
\]
Notice that $\xi_A=0$ if and only if $\mathbf{B}(\kappa_A)=0$. Thus,
one can write 
\[
\mathcal{D}_{(AB} \kappa_{C)} n^{AB} \hat{\pi}^C =  \big(
\mathbf{B}(\kappa_C)+\sqrt{2}\gamma_C{}^P \slashed{\mathcal{D}}^Q{}_P \kappa_Q \big)\hat{\pi}^C.
\]
A similar calculation as before shows that 
\[
n^{AB}\kappa^C \widehat{\mathcal{D}_{(AB} \pi_{C)}} = -\kappa^C
\big(\widehat{\mathbf{B}(\pi_C)} +\sqrt{2} \widehat{\gamma_C{}^P\slashed{\mathcal{D}}^Q{}_P\pi_Q}\big).
\]
Thus, one finds that
\begin{eqnarray*}
&&\hspace{-8mm} \int_{\mathcal{S}}\mathbf{L}(\kappa_A) \hat{\pi}^A \mathrm{d}\mu - \int_{\mathcal{S}}
\kappa_A \widehat{\mathbf{L}(\pi^A)} \mathrm{d}\mu \\
&& \hspace{1mm} =
\oint_{\partial\mathcal{S}}\left( \mathbf{B}(\kappa_A) \hat{\pi}^A -
   \kappa_A \widehat{\mathbf{B}(\pi^A)}\right) \mathrm{d}S + I,\\
\end{eqnarray*}
where
\[
I \equiv \sqrt{2}\oint_{\partial\mathcal{S}} \big(
   \gamma_C{}^P\slashed{\mathcal{D}}_P{}^Q \kappa_Q \hat{\pi}^C +
   \kappa^C
   \widehat{\gamma_C{}^P\slashed{\mathcal{D}}_P{}^Q\pi_Q}\big)\mbox{d}S. 
\]
In order to simplify the integral $I$ it is convenient to write the 2-dimensional Sen connection $\mathcal{D}_{AB}$ in terms of the Levi-Civita connection $D_{AB}$ as
\[
\slashed{\mathcal{D}}_{AB}\kappa_C = \slashed{D}_{AB} \kappa_C +
Q_{AB}{}^S{}_C\kappa_S, 
\]
where $Q_{AB}{}^S{}_C$ is the associated transition spinor between the connections. It follows then, after some calculations, that
\begin{eqnarray*}
&&\hspace{-6mm}  I = \sqrt{2}\oint_{\partial \mathcal{S}} \big( \gamma_C{}^P
   \slashed{D}_P{}^Q\kappa_Q \hat{\pi}^C + \kappa^C \gamma_C{}^P
   \slashed{D}_P{}^Q \hat{\pi}_Q \\
&& \hspace{6mm} +\gamma_C{}^P Q_P{}^{QS}{}_Q \kappa_S \hat{\pi}^C
   -\kappa^C \gamma_C{}^P \hat{Q}_P{}^{QS}{}_Q\hat{\pi}_S\big)
   \mathrm{d}S, \\
&& \hspace{-6mm}\phantom{I} = \sqrt{2}\oint_{\partial \mathcal{S}} \big(-\gamma_C{}^P
   \kappa_Q \slashed{D}_P{}^Q \hat{\pi}^C +
   \kappa^C\gamma_C{}^P\slashed{D}_P{}^Q\hat{\pi}_Q\\
&& \hspace{6mm} +\gamma_C{}^P Q_P{}^{QS}{}_Q \kappa_S \hat{\pi}^C
   -\kappa^C \gamma_C{}^P \hat{Q}_P{}^{QS}{}_Q\hat{\pi}_S\big)
   \mathrm{d}S,
\end{eqnarray*}
where in the second equality integration by parts on a manifold
without boundary has been used on the first integrand. Remarkably,
using the Jacobi identity for $\epsilon_{AB}$ one has that
\[
\gamma_C{}^P\kappa_Q\slashed{D}_P{}^Q\hat{\pi}^C = -\gamma_Q{}^P\kappa^Q\slashed{D}_{PC}\hat{\pi}^C,
\]
from where one concludes that 
\[
I =\sqrt{2}\oint_{\partial\mathcal{S}}\big( \gamma_C{}^P Q_P{}^{QS}{}_Q \kappa_S \hat{\pi}^C
   -\kappa^C \gamma_C{}^P \hat{Q}_P{}^{QS}{}_Q\hat{\pi}_S\big)
   \mathrm{d}S.
\]
Thus, the integrand in $I$ contains no differential operators acting
on $\kappa_C$ or $\hat{\pi}^C$. Accordingly, the boundary operator
$\mathbf{B}$ is, up to the vanishing of $I$, self-adjoint. Now, it can
be shown that, in fact, one has that 
\begin{equation}
Q_A{}^P{}_{BP} = -\hat{Q}_B{}^P{}_{AP} = \rho\, o_A \iota_B -\rho'
\iota_A o_B,
\label{TransitionSpinorHermitianConjugate}
\end{equation}
where the GHP coefficients $\rho$ and $\rho'$ are closely related to the expansions of the boundary $\partial\mathcal{S}$---see (\ref{thrho}). From the expression \eqref{TransitionSpinorHermitianConjugate} one readily concludes that $I =0$. Consequently, it follows that
\[
\mathbf{B}^*(\pi_A) = \mathbf{B}(\pi_A).
\]
Hence, we conclude that the boundary
operator $\mathbf{B}$ is self-adjoint.

\medskip
Substituting the Ansatz
\eqref{Ansatz:SolutionApproximateTwistor} into the approximate Killing
spinor equation \eqref{ApproximateTwistorEquation} one obtains the
following inhomogeneous equation for $\theta_A$:
\begin{equation}
\mathbf{L}(\theta_A )=F_A, \qquad F_A\equiv
-\mathbf{L}(\mathring{\kappa}_A). 
\label{InhomogeneousEquation}
\end{equation}
As by construction $\mathcal{D}_{(AB}\mathring{\kappa}_{C)}\in
H^\infty_{-3/2}$, one concludes that $F_A\in H^\infty_{-5/2}$. To
analyse the solvability of equation \eqref{InhomogeneousEquation} we
make use of a boundary value problem version of the Fredholm
alternative  adapted to weighted
Sobolev spaces---see e.g. \cite{Wlo87}.  More precisely, as $\mathbf{L}$ and $\mathbf{B}$ are self-adjoint, one
has that
\begin{equation}
\mathbf{L}(\theta_A) = F_A, \qquad \mbox{with}\quad
\mathbf{B}(\theta_A)|_{\partial\mathcal{S}} = G_A
\label{InhomogeneousSystem}
\end{equation}
has a solution if and only if
\[
\int_{\mathcal{S}} F_A\hat{\nu}^A \mbox{d}\mu +
\oint_{\partial\mathcal{S}} G_A \hat{\nu}^A \mbox{d}S =0,
\]
for all $\nu_A\in H^2_{-1/2}$ such that 
\begin{equation}
\mathbf{L}(\nu_A) =0, \qquad \mbox{with} \quad
\mathbf{B}(\nu_A)|_{\partial\mathcal{S}}=0.
\label{AdjointProblem}
\end{equation}
Thus, in the following we analyse the conditions under which the
adjoint problem \eqref{AdjointProblem} has a trivial Kernel. 

\smallskip
\noindent
\subsection{ Analysis of the Kernel of the adjoint problem} From the ellipticity of the operator $(\mathbf{L},\mathbf{B})$ it follows that the Kernel of the boundary value problem \eqref{AdjointProblem} is finite dimensional. Assume one has $\nu_A\in H^2_{-1/2}$ satisfying
\eqref{AdjointProblem}. Using integration by parts and the fall-off of $\nu_A$ it follows that 
\begin{eqnarray}
&& \int_{\mathcal{S}} \mathcal{D}_{(AB} \nu_{C)}
\widehat{\mathcal{D}^{AB}\nu^C}\mbox{d}\mu =  \oint_{\partial
  \mathcal{S}} n^{AB}\nu^C
   \widehat{\mathcal{D}_{(AB}\nu_{C)}}\mbox{d}S\nonumber\\
   &&
\phantom{\int_{\mathcal{S}} \mathcal{D}_{(AB} \nu_{C)}
\widehat{\mathcal{D}^{AB}\nu^C}\mbox{d}\mu} = \mathtt{H}[\nu_A,\bar\nu_{B'}]\geq 0
, \label{ObstructionKernel}
\end{eqnarray}
where following the discussion in the introduction we write
\[
\mathtt{H}[\nu_A,\bar\nu_{B'}]\equiv \oint_{\partial
  \mathcal{S}}\hat{\nu}^C \gamma_C{}^P  
\slashed{\mathcal{D}}^Q{}_P\nu_Q  \mbox{d}S\geq 0
\]
and to obtain the second equality we have used the identity
\begin{equation}
n^{AB}\hat{\nu}^C \mathcal{D}_{(AB} \nu_{C)} = \big(
\mathbf{B}(\nu_C) + \sqrt{2}\gamma_C{}^P \slashed{\mathcal{D}}^Q{}_P\nu_Q
\big) \hat{\nu}^C. 
\label{BoundaryIdentity}
\end{equation}
 Crucial in the sequel is that \emph{the eigenspinors of the 2-dimensional (Levi-Civita) Dirac operator $\slashed{D}{}^B{}_A\nu_B$ form a base of the space of smooth valence 1 spinors over $\partial \mathcal{S}$ which is orthonormal with respect to $L^2$ inner product induced by the Hermitian conjugation}---this follows from the ellipticity and self-adjointness of the operator---see e.g. \cite{Fri00,LawMic89,Eva98}. Now, if the Kernel of $(\mathbf{L},\mathbf{B})$ is non-trivial, it must contain spinors whose restriction to $\partial\mathcal{S}$ are  eigenspinors of the 2-dimensional Dirac operator. Now, if $\slashed{D}{}^B{}_A\nu_B=\lambda \nu_A$ then for a MOTS ($\rho=0$, $\rho'\geq 0$) a calculation readily gives that 
\begin{eqnarray*}
&& \mathtt{H}[\nu_A,\bar\nu_{B'}]= \lambda \oint_{\partial\mathcal{S}} \hat{\nu}^C\gamma_C{}^P \nu_P\mbox{d}S \\
&&\hspace{2.5cm}+ \frac{1}{2}\oint_{\partial\mathcal{S}} \rho'\big( \hat{\nu}^C \gamma_C{}^P\nu_P -\hat{\nu}^C\nu_C \big) \mathrm{d}S.
\end{eqnarray*}
The reality of $\mathtt{H}[\nu_A,\bar\nu_{B'}]$ and the fact that the eigenvalue $\lambda$ is purely imaginary (i.e. $\overline{\lambda} = - \lambda$) imply
\[
\lambda \oint_{\partial\mathcal{S}} \hat{\nu}^C\gamma_C{}^P \nu_P\mbox{d}S =0.
\]
From the latter and making use of the expansion $\nu_A = \nu_0 \iota_A - \nu_1 o_A$, one concludes that
\[
0\leq \mathtt{H}[\nu_A,\bar\nu_{B'}]= - \oint_{\partial\mathcal{S}} \rho'  |\nu_0|^2  \mathrm{d}S .
\]
This can only occur, for $\rho'>0$, if $\nu_A=0$ over $\partial\mathcal{S}$. It follows then from \eqref{ObstructionKernel} that if $\partial\mathcal{S}$ is a MOTS then  $\mathcal{D}_{(AB}\nu_{C)}=0$ on $\mathcal{S}$. That is, $\nu_A$ is a solution to
the spatial twistor equation that goes to zero at infinity. Using
Proposition 5 in \cite{BaeVal11b} then it follows that $\nu_A=0$ on
$\mathcal{S}$. This implies that there are no obstructions to the
existence of solutions to the system \eqref{InhomogeneousSystem}. The previous argument can be summarised in the following: 

\smallskip
\noindent
\textbf{Proposition.} 
{\em 
If $\rho'\geq 0$ and $\rho=0$ over $\partial\mathcal{S}$, then the boundary value problem 
 \[
\mathbf{L}(\kappa_A) =0, \qquad \mathbf{B}(\kappa_A)|_{\partial\mathcal{S}}= \sqrt{2} \gamma_A{}^P \hat{\phi}_P,
\]
with $\phi_A$ a smooth spinorial field over $\partial \mathcal{S}$
admits a unique solution of the form
 \eqref{Ansatz:SolutionApproximateTwistor}. Accordingly, there exists a spinor $\zeta_A$ such that in the
 asymptotic end it satisfies 
\[
 \zeta_\bmA = -\bigg( 1-\frac{m}{r}\bigg)\iota_\bmA + o_\infty(r^{-3/2}).
\]
}

\smallskip
\noindent
The above proposition holds even in the case that $\partial\mathcal{S}$ has several connected components each one being a MOTS
---that is, in the case  $(\mathcal{S}, h_{ij}, K_{ij})$ is a \emph{multiple back hole} initial data set.

\subsection{Main inequality in terms of boundary data} The right-hand side of the main inequality \eqref{BasicInequality} can be written in terms of the boundary data. The key observation is that the boundary condition $\hat{\xi}_A =\phi_A$ together with the the approximate twistor equation \eqref{ApproximateTwistorEquation} and its alternative form
\begin{equation} \label{ApproximateTwistorEquation2}
\mathcal{D}^{BC} \mathcal{D}_{BC} \kappa_A + \Omega_{ABCD} \mathcal{D}^{BC} \kappa^D + \frac{1}{3} K \mathcal{D}_{AB} \kappa^B = 0
\end{equation}
allow to systematically eliminate all the transverse derivatives $\slashed{\mathcal{D}}\kappa_C$ in the integral over $\partial\mathcal{S}$. We can write the right-hand side of main inequality as
\begin{eqnarray*} 
&& \frac{1}{\sqrt{2}}\oint_{\partial \mathcal{S} } \gamma_{AB} \zeta_C \widehat{\mathcal{D}^{(AB} \zeta^{C)}} \mbox{d} S \\ 
&& \hspace{1cm}=\frac{1}{\sqrt{2}}\oint_{\partial \mathcal{S} } \gamma^{AB} \phi^C \left(\hat{\phi}^D \Omega_{ABCD} + \tfrac{2}{3} \slashed{\mathcal{D}}_{AC} \hat{\phi}_B \right) \mbox{d} S \\
&& \hspace{1.5cm}- \frac{\sqrt{2}}{3} \oint_{\partial \mathcal{S} } \phi^C \slashed{\mathcal{D}} \xi_C \mbox{d} S.
\end{eqnarray*}
The alternative form of the approximate twistor equation given by equation  \eqref{ApproximateTwistorEquation2} yields
\begin{equation} \nonumber
\begin{split}
 \mathcal{D}^{BC} \xi_{BCA} & - \slashed{\mathcal{D}}_A{}^B \xi_B + \frac{1}{2} \gamma_{A}{}^B \slashed{\mathcal{D}} \xi_B + \frac{1}{2} K \hat{\phi}_A \\
& + \Omega_{ABCD} \mathcal{D}^{BC}\kappa^D = 0,
\end{split}
\end{equation}
but from approximate twistor equation the first and the last terms cancels each other out, so that
\[
\slashed{\mathcal{D}} \xi_C = - 2 \gamma_{C}{}^A \slashed{\mathcal{D}}_A{}^B \hat{\phi}_B - K \gamma_C{}^A \hat{\phi}_A.
\]
After performing integration by parts, the main inequality \eqref{BasicInequality} reads
\[
 4\pi m \geq \sqrt{2} \oint_{\partial\mathcal{S}}\hat{\phi}^A \gamma_A{}^B  \slashed{\mathcal{D}}_{BC} \phi^C \mathrm{d}S.
\]

\section{Mass inequalities}
We are now ready to state the main result of this paper. Given a hypersurface $\mathcal{S}$ and smooth spinor $\phi_A$ defined over a MOTS $\partial\mathcal{S}$ one has that
\begin{equation}
4\pi m \geq \frac{\kappa}{\sqrt{2}} \mathtt{H}[\phi_A,\bar\phi_{A'}],
\label{MainInequality}
\end{equation}
where
\[
\mathtt{H}[\phi_A,\bar{\phi}_{A'}]\equiv \frac{2}{\kappa} \oint_{\partial\mathcal{S}}\hat{\phi}^A \gamma_A{}^B  \slashed{\mathcal{D}}_{BC} \phi^C \mathrm{d}S. 
\]

\smallskip
\noindent
Given two spinors $\kappa_A$ and $\omega_{A}$, the functional
$\mathtt{H}[\kappa_A,\omega_{B}]$ coincides with the
\emph{Nester-Witten functional}---see
e.g. \cite{HorTod82,ReuTod84,Sza08}---which plays a role in various
quasilocal energy constructions.  \emph{If the spinor $\phi_A$ could
be chosen in such a way that $\mathtt{H}[\kappa_A,\omega_{B}]$ is
manifestly non-negative, one would have obtained a non-trivial bound
on the ADM mass of the black hole. Consequently, inequality
\eqref{MainInequality} can be used as the starting point for the
construction of new geometric inequalities involving the mass.} As
examples of interesting choices of $\phi_A$ consider:
 
 \medskip
 \noindent
 \emph{(i)} The simple choice $\phi_A=0$ over $\partial \mathcal{S}$ leads to \emph{a new proof of the positivity of the mass of a black hole}, i.e. $m\geq 0$. 
 
 \medskip
 \noindent
 \emph{(ii)}  Choosing $\phi_A$ to be an eigenspinor of the 2-dimensional Dirac operator, i.e. $\slashed{D}_A{}^B \phi_B = \lambda \phi_A$, it follows from the fact that the eigenvalue must be pure imaginary, i.e. $\overline{\lambda} = - \lambda$, and the reality of $\mathtt{H}[\phi_A,\bar\phi_{A'}]$ that 
    \begin{equation} \label {boundphi}
    \oint_{\partial\mathcal{S}} |\phi_0|^2 \mathrm{d}S = \oint_{\partial\mathcal{S}} |\phi_1|^2\mathrm{d}S.
    \end{equation}
    Moreover, inequality \eqref{MainInequality} takes the form
    \begin{equation}
    4 \pi m  \geq  \sqrt{2} \oint_{\partial\mathcal{S}} \rho' |\phi_0|^2 \mathrm{d}S.
    \label{DiracIneqIntermmediate}
    \end{equation}
    Now, on generic topological spheres the eigenspace associated to a given eigenvalue is 2-dimensional. The pair $\{\phi_A,\hat{\phi}_A \}$ can be shown to be a basis of the eigenspace and  to be non-zero everywhere on $\partial\mathcal{S}$---see e.g. \cite{Her12}, Theorems 6.2.5 and 6.2.6. Now, choosing the (pointwise) normalisation $\phi_A\hat{\phi}{}^A=1$, it readily follows from \eqref{boundphi} that 
    \[
    \oint_{\partial\mathcal{S}} |\phi_0|^2 \mathrm{d}S=\tfrac{1}{2}|\partial\mathcal{S}|,
    \]
    where $|\partial\mathcal{S}|$ denotes the area of $\partial \mathcal{S}$. Combining this last observation with inequality \eqref{DiracIneqIntermmediate} one concludes that
    \[
    4\pi m \geq \frac{\sqrt{2}}{2} (\min_{\partial\mathcal{S}}\rho')\, |\partial \mathcal{S}| .
    \]
It is worth to notice that for a MOTS $\rho'$ coincides with the mean curvature $h$ of the $\partial S$, such that this inequality is equivalent with
    \[
    4\pi m \geq \frac{\sqrt{2}}{2} (\min_{\partial\mathcal{S}}h)\, |\partial \mathcal{S}| .
    \]
To the author's best knowledge, this inequality is new.

 \medskip
 \noindent
 \emph{(iii)} \emph{Relation to the area variation} \cite{Tod91}. Let $H_{AA'} = \rho \iota_A\bar\iota_{A'} + \rho' o_A \bar{o}_{A'}$
     denote the spinorial counterpart of the mean curvature vector to $\partial\mathcal{S}$. The variation of the area $|\partial \mathcal{S}|$ in the direction of a vector $v^a$ on $\mathcal{M}$ is given by the formula
     \[
     \delta_v |\partial\mathcal{S}| = -\oint_{\partial\mathcal{S}} v^{AA'} H_{AA'}\mbox{d}S,
     \]
     where $v^{AA'}$ is the spinorial counterpart of $v^a$. In the space-spinor formalism the mean curvature vector reads
\[     
H_{AB} = - \rho \iota_{A} o_{B} + \rho' o_{A} \iota_B.
\]
 Making the choice $\phi_A= - \phi_0 \iota_A$ (i.e. $\phi_1=0$) one
 then has that 
\[
v_{AB}\equiv -\phi_{(A} \hat{\phi}_{B)}=\frac{1}{2}|\phi_0|^2
\gamma_{AB}
\]
 can be interpreted as the spinorial counterpart of the
(outwardpointing) radial vector to $\partial\mathcal{S}$.  For this
choice the right-hand side of \eqref{MainInequality} for a MOTS can be
written in terms of a variation of its area with respect to flow
generated by $v^a$. More precisely, one has that
     \[
     4 \pi m \geq  2 \sqrt{2} \delta_v |\partial S|.
     \]

\medskip
For the sake of simplicity, the above statements have been formulated
for $\partial\mathcal{S}$ consisting of a single connected
component. However, the methods presented here also applies to an
inner boundary consisting of several components, each one with the
topology of $\mathbb{S}^2$ and satisfying the MOTS condition.

\section{Conclusions}
In this article we have developed a new strategy for the construction
of geometric inequalities involving the ADM mass of a black hole
spacetime. This approach relies heavily on the use of spinors and has
the remarkable property of allowing the specification of the two
components of a valence-1 spinor $\phi_A$ defined over a MOTS. The use
of the MOTS condition is central in the solvability of the boundary
value problem for the approxmate twistor equation. However,
it is not necessary in the argument showing that the
rigth-hand side of inequality \eqref{MainInequality} can be expressed
purely in terms of boundary data. 

The main question is whether the methods developed in this article can
be used to make inroads towards a general proof of Penrose's
inequality. In \cite{Her97b} Witten's approach to the positivity of
the mass was used to obtain a Penrose-like inequality---i.e. an
inequality involving the ADM mass and the square root of the area
which, in addition, contains further constant which is hard to control
given the rigidity in the specification of boundary data. The main
idea in that article was to study the change of the mass under
conformal rescalings of the 3-metric. A similar strategy can be
followed with the framework presented in the present article. The
further flexibility given by the possibility of prescribing full
boundary data could prove crucial in controlling constants appearing
in the analysis. 

Finally, it is pointed out that it would also be interesting to
analyse whether the methods in this article can be adapted to settings
with different asymptotic boundary conditions---e.g. hyperboloidal
ones so that a connection with the Bondi mass can be established. 

\medskip
The ideas expressed in the previous paragraphs will be pursued elsewhere.

\section*{Acknowledgements}
JAVK is grateful to Jos\'{e} Luis Jaramillo, Laszlo B. Szabados,
Thomas B\"ackdahl, Mahdi Godazgar and Bernardo Araneda for many
stimulating discussions on the topic of this paper. JK would like to
acknowledge networking support by the COST Action GWverse CA16104 and the support of Center for Theoretical Physics PAS. We
further thank Thomas B\"ackdahl for his advice on the use of {\tt xAct}.

\section{Appendix}
\subsection{Irreducible decompositions} 
Given a spinor $\kappa_A$ define
\[
\xi_A \equiv \tfrac{2}{3}\mathcal{D}_A{}^Q\kappa_Q, \qquad \xi_{ABC}
  \equiv \mathcal{D}_{(AB} \kappa_{C)}. \label{DKappaDecomposition}\\
\]
One then has the decomposition
\[
\mathcal{D}_{AB} \kappa_C = \xi_{ABC} - \xi_{(A} \epsilon_{B)C}.
\]
\subsection{Integration by parts}
\label{Appendix:IntegrationByParts}

Integration by parts on the 3-manifold $\mathcal{S}$ with respect to the Sen connection $\mathcal{D}_{AB}$ is carried out according to the identity \begin{eqnarray*}
&& \int_{\mathcal{U}} \mathcal{D}_{AB} \kappa_C \hat{\zeta}^{ABC}
\mbox{d}\mu = \oint_{\partial \mathcal{U}} \widetilde{n}_{AB} \kappa_C
\hat{\zeta}^{ABC} \mbox{d}S \nonumber \\
&& \hspace{1cm} + \int_{\mathcal{U}} \kappa^C \big(
\widehat{\Omega_C{}^{ABD}\zeta_{ABD}} - \widehat{\mathcal{D}^{AB}
  \zeta_{ABC}} \big) \mbox{d}\mu, 
\end{eqnarray*}
with $\mathcal{U}\subset \mathcal{S}$ and where $\mbox{d}S$ denotes
the area element of $\partial \mathcal{U}$, $\widetilde{n}_{AB}$ its outward pointing ("outside" of $\ \mathcal{U}$) normal and $\zeta_{ABC}$ is an arbitrary symmetric spinor.

Integration by parts on $\partial\mathcal{S}$ proceeds in the same lines as on $\mathcal{S}$ with the added simplification of not giving rise to boundary terms. Thus, for symmetric spinors $\kappa_A$ and $\zeta_{ABC}$ one has that 
\[
\oint_{\partial\mathcal{S}} \slashed{D}_{AB}\kappa_C \zeta^{ABC}\mathrm{d}S = - \oint_{\partial\mathcal{S}
}\kappa_C \slashed{D}_{AB}\zeta^{ABC} \mathrm{d}S.
\]
In some cases it is necessary to use integration by parts on expressions involving components. The following identities have been proven in \cite{PenRin84}:
\[
\oint_{\partial\mathcal{S}} \chi \eth \eta \mbox{d}S = - \oint_{\partial\mathcal{S}}\eta \eth \chi \mbox{d}S
\label{IntegrationByPartsEth}
\]
if the GHP types of $\chi$ and $\eta$ add up to $\{-1,1\}$, and 
\[
\oint_{\partial\mathcal{S}} \chi \eth' \eta \mbox{d}S = - \oint_{\partial\mathcal{S}}\eta \eth' \chi \mbox{d}S
\label{IntegrationByPartsEthDash}
\]
if the type of $\chi$ and $\eta$ add up to $\{1,-1\}$.
\subsection{Commutators}
Several of the calculations require the
commutators between the various covariant derivatives. The commutator
between the 3-dimensional Sen connection on an hypersurface,
assuming the vacuum Einstein field equations hold, can
be expressed as
\begin{eqnarray*}
&& [\mathcal{D}_{AB}, \mathcal{D}_{CD}]\kappa_E=\tfrac{1}{2}\big(
   \epsilon_{A(C}\square_{D)B} + \epsilon_{B(C}\square_{D)A}
   \big)\kappa_E\\
&& \hspace{1cm} + K_{CDQ(A}\mathcal{D}_{B)}{}^Q\kappa_E - K_{ABQ(C}\mathcal{D}_{D)}{}^Q\kappa_E,
\end{eqnarray*}
see e.g. \cite{BaeVal11a}, 
where $\square_{AB}$ denotes the usual Penrose box---see
\cite{PenRin84}. Now, using the above commutator one can write
\[
\mathcal{D}_{AQ}\mathcal{D}_B{}^Q =
\tfrac{1}{2}\epsilon_{AB}\mathcal{D}_{PQ}\mathcal{D}^{PQ} + \Delta_{AB}
\]
where 
\[
\Delta_{AB} \equiv \mathcal{D}_{C(A}\mathcal{D}_{B)}{}^C.
\]
A calculation using the expression for $[\mathcal{D}_{AB},
\mathcal{D}_{CD}]$ readily yields that
\[
\Delta_{AB}\kappa_C = \square_{AB}\kappa_C
-K_{APQB}\mathcal{D}^{PQ}\kappa_C -K_{P(A|Q|}{}^P\mathcal{D}_{B)}{}^Q\kappa_C.
\]
One can rewrite the action of $\Delta_{AB}$ as
\[
\Delta_{AB} = \tfrac{1}{2}\square_{AB} - \tfrac{1}{2} \Omega_{ABPQ}\mathcal{D}^{PQ} + \tfrac{1}{3}K\mathcal{D}_{AB}.
\]
Similarly, for the 2-dimensional Sen connection one can define
\[
\slashed\Delta \equiv \slashed{\mathcal{D}}_{AB}\slashed{\mathcal{D}}^{AB}, \qquad \slashed{\Delta}_{AB}\equiv \slashed{\mathcal{D}}_{C(A}\slashed{\mathcal{D}}_{B)}{}^C.
\]
In particular, we have that
\[
\slashed{\mathcal{D}}_{CA}\slashed{\mathcal{D}}_{B}{}^C= \tfrac{1}{2}\epsilon_{AB}\slashed\Delta + \slashed{\Delta}_{AB}.
\label{DecompositionCommutatorSen2D}
\]

\subsection{The Lopatinskij-Shapiro conditions}

To establish the compatibility of the approximate twistor
equation and the transverse boundary conditionnone needs to consider the so-called
\emph{Lopatinskij-Shapiro conditions}---see
e.g. \cite{Dai06,WloRowLaw95}. Using the decomposition of $\mathcal{D}_{AB}$ in terms of $\slashed{\mathcal{D}}$ and $\slashed{\mathcal{D}}_{AB}$, the
principal part of the approximate twistor equation takes the form
\begin{equation}
\mathcal{D}^{PQ}\mathcal{D}_{PQ}\kappa_A=\slashed{\mathcal{D}}^{PQ}\slashed{\mathcal{D}}_{PQ}\kappa_A -\tfrac{1}{2}\slashed{\mathcal{D}}^2 \kappa_A,
\label{PrincipalPartApproximateTwistor}
\end{equation}
while for the transverse  boundary condition one gets
\begin{equation}
\mathcal{D}^P{}_A \kappa_P = \slashed{\mathcal{D}}^P{}_A\kappa_P- \tfrac{1}{2}\gamma^P{}_A \slashed{\mathcal{D}}\kappa_P .
\label{PrincipalPartBoundaryCondition}
\end{equation}
In a neighbourhood of $\partial\mathcal{S}$ one chooses coordinates so
that the location of the boundary is given by the condition $\rho=0$
and $\slashed{\mathcal{D}}=\partial_\rho$. To verify the
Lopatinskij-Shapiro conditions one considers decaying solutions to the
auxiliary ordinary differential equations problem
\begin{subequations}
\begin{eqnarray}
&& \kappa''_A -2|\xi|^2 \kappa_A =0,  \label{LS1}\\
&&\big( \gamma^P{}_A \kappa'_P -2 \mbox{i}\xi^P{}_A
   \kappa_P\big)\big|_{\rho=0}=0, \label{LS2}
\end{eqnarray}
\end{subequations}
obtained from the principal parts \eqref{PrincipalPartApproximateTwistor} and 
 \eqref{PrincipalPartBoundaryCondition} by
the replacements $\slashed{\mathcal{D}}\mapsto '$,
$\slashed{\mathcal{D}}_{AB}\mapsto \mbox{i}\xi_{AB}$ where
$\xi_{AB}=\xi_{(AB)}$ is an arbitrary non-zero real rank 2 spinor
---i.e. $\hat{\xi}_{AB}=-\xi_{AB}$,
$|\xi|^2\equiv\xi_{PQ}\hat{\xi}^{PQ}$, $\gamma^{AB}\xi_{AB}=0$. Moreover, $'$ denotes differentiation
with respect to $\rho$. The decaying solutions of equation \eqref{LS1}
are given by
\[
\kappa_A = \kappa_{A\star} e^{-|\xi|^2\rho},
\]
 where $\kappa_{A\star}$ is constant. Substitution of the latter
into equation \eqref{LS2} leads to the condition
\[
(2\mbox{i}\xi^P{}_A + \gamma^P{}_A|\xi|^2)\kappa_{P\star}=0 \qquad
\mbox{on} \qquad \partial\mathcal{S},
\]
from which, taking into account that both $\xi_{AB}$ and $\gamma_{AB}$ are
real spinors, it follows that $\kappa_{A\star}=0$. Thus, the approximate twistor equation
 with the transverse boundary condition
 satisfies the Lopatinskij-Shapiro
condition, so the associated boundary value problem is elliptic.

\subsection{Proofs of various properties of the Kernel of the adjoint
  problem}
  
\subsubsection{The Kernel of $(\mathbf{L},\mathbf{B})$ includes the Kernel of $\mathbf{D}$}  
  
To show that an element of the kernel of the
adjoint problem is also a solution to the Sen-Witten-Dirac equation one starts by considering the $L^2$-norm of the Sen-Witten-Dirac operator acting on the element of the Kernel of $(\mathbf{L},\mathbf{B})$. Then, using integration by parts
it follows that 
\begin{eqnarray*}
&& 0\leq \int_{\mathcal{S}} \mathcal{D}_A{}^B\nu_B
   \widehat{\mathcal{D}^{AC}\nu_C} \mbox{d}\mu\\
&& \phantom{0}= \oint_{\partial\mathcal{S}} \hat{\nu}_C
   n^{AC}\mathcal{D}_A{}^B\nu_B \mbox{d}S - \oint_{\partial\mathcal{S}_\infty} \hat{\nu}_C
   n^{AC}\mathcal{D}_A{}^B\nu_B \mbox{d}S \\
&& \hspace{1cm} + \int_{\mathcal{S}} \bigg(
   \hat{\nu}_C\mathcal{D}^{AC} \mathcal{D}_A{}^B \nu_B-\tfrac{1}{2}K \hat{\nu}^A \mathcal{D}_A{}^B\nu_B
   \bigg) \mbox{d}\mu.\\
\end{eqnarray*}
Now, the boundary integral at $\partial\mathcal{S}$ vanishes as a
consequence of $\mathbf{B}(\nu_A)=0$ while that
at the sphere at infinity also vanishes as
$\hat{\nu}_A\mathcal{D}^{AC}\nu_C=o(r^{-2})$ in the asymptotic
end. Now, making use of the decomposition
\[
\mathcal{D}_{AC}\mathcal{D}_B{}^A =\tfrac{1}{2}\epsilon_{CB}\Delta +
\Delta_{CB}
\]
one has, further, that 
\begin{eqnarray*}
&& \int_{\mathcal{S}} \mathcal{D}_A{}^B\nu_B
   \widehat{\mathcal{D}^{AC}\nu_C} \mbox{d}\mu=  \tfrac{1}{2}\int_{\mathcal{S}} \hat{\nu}^C \Delta\nu_C
   \mbox{d}\mu  \\
&& \hspace{1cm} - \int_{\mathcal{S}} \hat{\nu}^C \Delta_{CB} \nu^B \mbox{d}\mu+
   \tfrac{1}{2}\int_{\mathcal{S}} K \hat{\nu}_C \mathcal{D}^{CB}\nu_B \mbox{d}\mu.
\end{eqnarray*}
Observing that in vacuum one has 
\[
\Delta_{CB} \nu^B=-\frac{1}{2}\Omega_{CBAD}\mathcal{D}^{AD}\nu^B
+\frac{K}{3}\mathcal{D}_{CB}\nu^B
\]
 and using the expression for
$\Delta \nu_C \equiv \mathcal{D}_{AB} \mathcal{D}^{AB} \nu_C$
given by the approximate Killing spinor equation one concludes that the
right-hand side of the last equality vanishes and thus
\[
\int_{\mathcal{S}} \mathcal{D}_A{}^B\nu_B
   \widehat{\mathcal{D}^{AC}\nu_C} \mbox{d}\mu=0
\]
so that $\mathcal{D}_A{}^B\nu_B=0$ on $\mathcal{S}$. 

\subsection{Properties of the 2-dimensional Sen-Witten-Dirac operator}
\noindent A calculation readily shows that in GHP notation the equation
$\slashed{\mathcal{D}}_A{}^B\nu_B=0$ implies that
\begin{eqnarray*}
&& \eth' \nu_0 +\rho\, \nu_1=0, \\
&& \eth \nu_1 + \rho'\, \nu_0=0.
\end{eqnarray*}
Using the methods of the Appendix in \cite{Sza94b} one can show that if either $\rho=0$ or $\rho'=0$ then necessarily $\nu_0=\nu_1=0$
so that $\nu_A=0$---that is, the Kernel of
$\slashed{\mathcal{D}}_A{}^B\nu_B$ is trivial. 

Now, a computation readily shows that 
\[
\oint_{\partial\mathcal{S}} \slashed{\mathcal{D}}_A{}^B\kappa_B \mathrm{d}S = \oint_{\partial \mathcal{S}}\big( \widehat{\slashed{\mathcal{D}}^B{}_A\pi_B}   - 2 Q_B{}^C{}_{AC}\hat{\pi}^B\big)\kappa^A\mathrm{d}S
\]
so that $\slashed{\mathcal{D}}_A{}^B\nu_B$ is not self-adjoint unless $\rho=\rho'=0$. Expanding the adjoint operator
\[
    \slashed{\mathcal{D}}_A{}^B\pi_B -2 Q_A{}^{CB}{}_C\pi_B
\]
in terms of a dyad yields the components
\begin{eqnarray*}
&& \eth'\pi_0-\rho\pi_1, \\
&& \eth\pi_1-\rho'\pi_0.
\end{eqnarray*}

\medskip
Of particular interest in the present analysis is the eigenvaule problem for the 2-dimensional Sen-Witten-Dirac operator---i.e.
\[
\slashed{\mathcal{D}}_A{}^B\kappa_B =\lambda\kappa_A.
\]
Applying the operator once more and integrating gives
\[
\oint_{\partial\mathcal{S}} \hat{\kappa}^C
\slashed{\mathcal{D}}_C{}^B\slashed{\mathcal{D}}^A{}_B\kappa_A \mbox{d}S=\lambda^2 \oint_{\partial\mathcal{S}}\kappa_C\hat{\kappa}^C\mbox{d}S.
\]
Integration by parts plus some further manipulations eventually leads to
\begin{eqnarray*}
&& 0\leq \oint_{\partial\mathcal{S}} \big( \slashed{\mathcal{D}}^P{}_A \kappa_P  \big)\widehat{\big(\slashed{\mathcal{D}}^{QA}\kappa_Q  \big)}\mbox{d}S\\
&& \hspace{1cm}=-\lambda^2 \oint_{\partial\mathcal{S}} (|\kappa_0|^2 +|\kappa_1|^2)\mbox{d}S\\
&& \hspace{1.5cm}+2 \lambda \oint_{\partial\mathcal{S}} ( \rho |\kappa_0|^2 +\rho'|\kappa_1|^2)\mbox{d}S. 
\end{eqnarray*}
From the above inequality it follows the (classic) observation that if $\rho=\rho'=0$ then the eigenvalues of the Dirac operator are pure imaginary. If, for example, $\rho=0$ and $\rho'>0$ (MOTS) then this is no longer true a the eigenvalues are general complex numbers.

\subsection{Nester--Witten functional}
\emph{Sparling's form} is defined as 
\[
\mathbf{\Gamma}(\lambda_A,\bar\mu_{B'}) \equiv \mbox{i} \nabla_{\bmB\bmB'}\lambda_\bmA \nabla_{\bmC\bmC'}\bar\mu_{\bmA'}\mathrm{d}x^{\bmA\bmA'}\wedge \mathrm{d}x^{\bmB\bmB'}\wedge \mathrm{d}x^{\bmC\bmC'}.  
\]
It is Hermitian in the sense that 
\[
\overline{\mathbf{\Gamma}(\lambda_A,\bar\mu_{B'})}=\mathbf{\Gamma}(\lambda_A,\bar\mu_{B'}).
\]
In vacuum Sparling's form is exact---i.e. $\mathrm{d} \mathbf{u}= \Gamma$
for some 2-form $\mathbf{u}=u_{\mu\nu}\mathrm{d}x^\mu\wedge \mathrm{d}x^\nu$. This 2-form is used, in turn, to define the \emph{Nester-Witten functional} over a 2-surface $\partial\mathcal{S}$ via
\[
\mathtt{H}[\lambda_R,\bar{\mu}_{S'}] \equiv \frac{2}{\kappa}\oint_{\partial\mathcal{S}} u_{\mu\nu}({\bm \lambda}, \bar{\bm \mu})\mathrm{d}x^\mu\wedge \mathrm{d}x^\nu.
\]
In \cite{Sza94b} it has been shown that the above functional can be rewritten as
\[
\mathtt{H}[\lambda_R,\bar{\mu}_{S'}]=\frac{2}{\kappa}\oint_{\partial\mathcal{S}} \bar\gamma^{R'S'}\bar\mu_{R'} \slashed{\mathcal{D}}^S{}_{S'}\lambda_S\mathrm{d}S,
\]
A calculation shows that, in terms of $SU(2,\mathbb{C})$  (i.e. space spinors), the above expression is equivalent to
\[
\mathtt{H}[\lambda_R,\bar{\mu}_{S'}]=\frac{2}{\kappa}\oint_{\partial\mathcal{S}} \hat{\gamma}_R{}^S\hat{\phi}{}^R\slashed{\mathcal{D}}_{P}{}^S\phi_S \mathrm{d}S.
\]

\end{document}